\documentclass[prb,showpacs, amsmath,amsfonts,amssymb,twocolumn,superscriptaddress]{revtex4} 
\usepackage{color,graphicx}
\usepackage{amsfonts}
\usepackage{bm}

\newcommand{\ET}{$\kappa$-(BEDT-TTF)$_2$Cu(NCS)$_2$}
\newcommand{\hc}{\text{H}_{c2}}

\begin{document}

\title{Critical field and Shubnikov-de Haas oscillations of
\ET under pressure}

\author{C. Martin}
\author{C. C. Agosta}
\email{cagosta@clarku.edu}
\affiliation{Department of Physics, Clark University, 950 Main St., Worcester, MA 01610}
\author{ S. W. Tozer}
\author{ H. A. Radovan}
\affiliation{The National High Magnetic Field Laboratory,
1800 E. Paul Dirac Dr., Tallahassee, FLÊ 32310}
\author{Tatsue Kinoshota}
\affiliation{CREST, JST, Kawaguchi 332-0012 Japan}
\author{M. Tokumoto}
\affiliation{Nanotechnology Research Institute, AIST, Tsukuba 305-8568, Japan}
\affiliation{CREST, JST, Kawaguchi 332-0012 Japan}

\date{\today}

\begin{abstract}
A tuned tank circuit in combination with a nonmetallic diamond anvil cell has been successfully used to measure the change in critical field with angle in \ET\ at pressures up to 1.75 kbar and at temperatures down to 70 mK.  The critical field has been found to decrease by more than 90\% within less than  2 kbar and at a much higher rate for the field applied parallel to the conducting planes.  For this orientation, at 1.75 kbar,  we have seen a clear change from the ambient pressure behavior of the critical field with temperature at low temperatures. Up to P = 1.75 kbar, the $\hc$($\theta $) phase diagram is in good agreement with the theoretical prediction for weakly coupled layered superconductors. We have also succeeded in measuring oscillations in the resistivity of the normal state at higher magnetic field. The $\alpha$-orbit Shubnikov-de Haas frequency was found to increase at a rate of 44 T/kbar. Our experiment opens the possibility for further investigations of the effective mass with pressure, especially because the setup is suitable for pulse fields as well.
\end{abstract}


\maketitle
\section{Introduction}

Due to its strong two-dimensional character, the charge-transfer
organic salt \ET\ is a suitable material for studying the various
theories put forth for anisotropic superconductivity in magnetic
fields. The electronic structure of organic superconductors is
very similar to that of the cuprate high $T_c$ superconductors,
consisting of stacks of alternating conducting and insulating
sheets. In contrast to the cuprates, however, the critical fields
of organic superconductors are much lower making them easier to
study. Among the organics, \ET\, with a $T_c = 10.4 K$, has been
shown to be one of the compounds with the highest critical
fields. Even with its conducting planes parallel to the applied
field, $\hc$ is less than 40 tesla (T).\cite{Ishiguro,Williams}
Furthermore, high-purity single crystals of \ET\ are available
which make for reliable studies of the Fermi surface.  For
example, the samples used in this study have mean free paths from
600-900 \AA,  ~and a superconducting coherence length in the layers
of $\sim$ 100 \AA.  These parameters put \ET\ clearly in the clean
limit. YBCO in comparison has a mean free path less than 100 \AA\
and ~and a superconducting coherence length in the layers of
$\sim$ 50 \AA.\cite{kresin}

\begin{figure}[h!] \begin{center}
\includegraphics[keepaspectratio=1,width=8 cm]{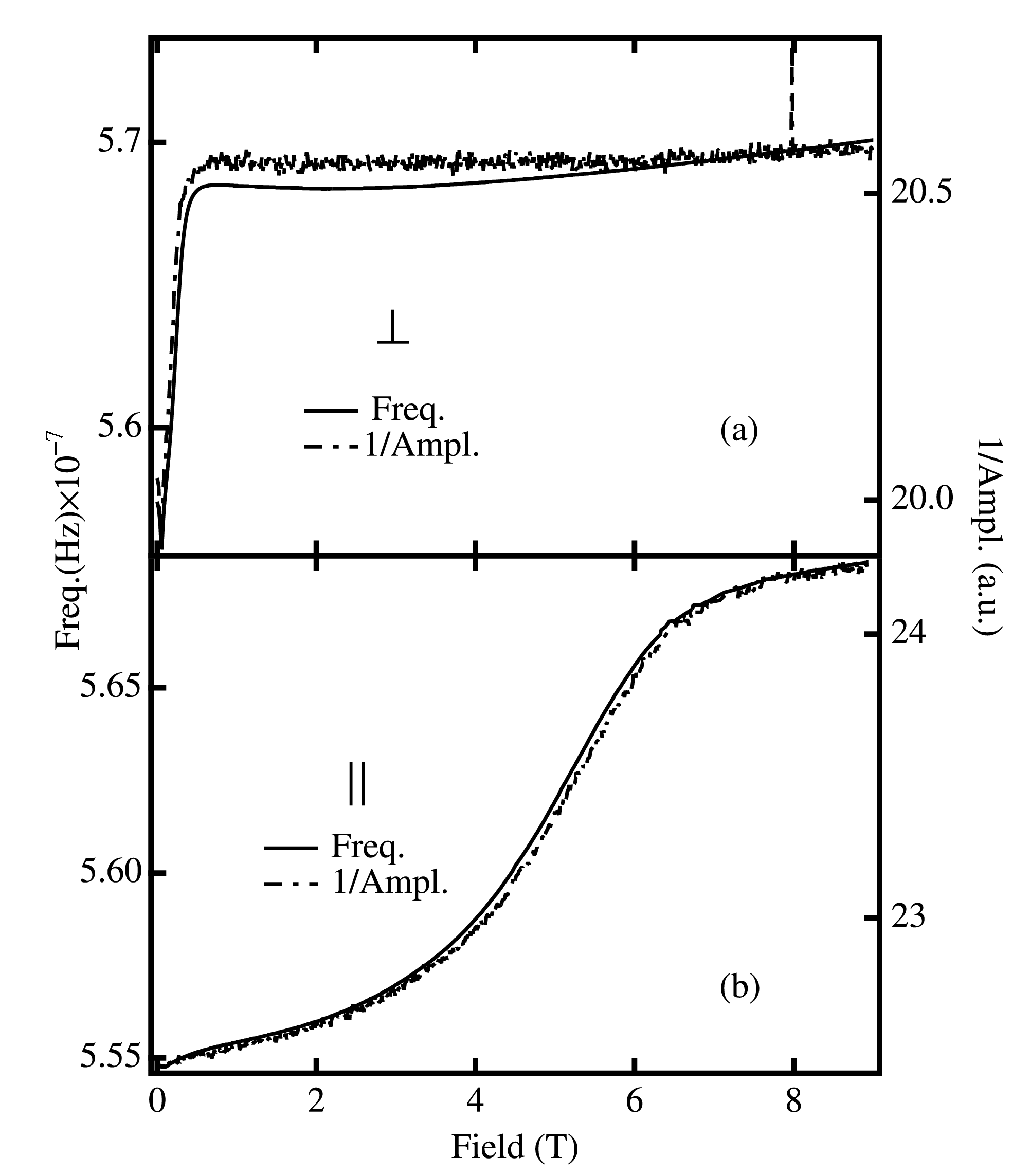}
\caption{\label{fig:ParPen} The rf penetration (proportional to the change in frequency)
as a function of magnetic field for the orientation perpendicular to the conducting
planes (a) and the parallel orientation (b). $\hc$ is determined by the intersecting point
between the linear extrapolation of the normal state and the superconducting
transition.}
\end{center} \end{figure}

The strong effect of the pressure on the band structure via
modification of the carrier effective mass and Fermi surface (and
hence on the superconducting properties) already reported in the
literature \cite{Caulfield} has motivated our work. Our innovation
was to combine a nonmetallic diamond anvil cell (DAC) \cite{tozer}
with an rf penetration depth technique. \cite{Coffey} The plastic
pressure cell design overcomes the difficulties of using metals in
magnetic field, can be made of a relatively small size to fit on a
rotating platform and, by placing a ruby chip inside the cell, the
pressure can be measured in-situ. The penetration depth was measured using the TDO technique, which offers the
advantage of not requiring contacts on the sample, and therefore,
eliminates problems like contact resistance and additional stress
on the sample. It is particularly well suited for use in the diamond anvil
cell because the coil and the sample can be of arbitrarily small
size. In a recent advance, we have succeeded in using this
combination of techniques in the pulsed field environment to 50 T
at He-3 temperatures.

In the present paper, we focus on the change in critical field under pressure for
different orientations of the applied dc magnetic field with
respect to the conducting planes (we will refer to $\theta $ as
the angle between the magnetic field and the normal to the
conducting planes). The study of reduced dimensional systems is important, because of
the different mechanisms which destroy the superconductivity when
the magnetic field is applied perpendicular or parallel to the
conducting layers.\cite{Ishiguro2}  In the perpendicular orientation,
or whenever the magnetic field has a nonzero component in this
direction, the upper critical field is determined by orbital
magnetic effects (through the in-plane kinetic energy of
electrons). However, when the applied magnetic field is parallel
or very close to the parallel direction, the orbital effects can
be suppressed, because the vortices can fit outside the conducting
planes. In this case, the spin-magnetic field interaction may
become important. In the absence of any other mechanisms
(spin-orbit scattering, many body effects), the maximum critical
field in the parallel orientation, called the BCS Pauli
paramagnetic limit,  $H_P^{BCS}$, is driven by the spin
polarization effect, where the condensation energy is overcome by
the Zeeman splitting energy.\cite{Clogston,Chandrasekhar}  For a
comprehensive summary of organic superconductors in the Pauli
paramagnetic limit region we refer the reader to Ref.
\onlinecite{Ishiguro}.

In particular, \ET\ has been found to have an unusual evolution of
the parallel critical field with temperature. In spite of their
differences in the measured values of the critical field and in
the cuvature of the phase diagram at higher temperatures, all the
experimental results at ambient pressure agree that the absolute
value of $\hc^\|$ not only exceeds $H_P^{BCS}$, but it also shows
no tendency of saturation at low temperatures \cite{Singleton,
zuo, Bayindir2} and displays  positive
curvature toward 0 K. However, the reason for this behavior is
not well understood. We cite two recent hypotheses, one that
explains the lack of saturation as a first order phase transition
into the FFLO state\cite{Singleton} and another one that claims
the high critical fields are indeed beyond the BCS Pauli
paramagnetic limit, but comparable to the paramagnetic limit
calculated from thermodynamic quantities.\cite{zuo}  Both
references seem to ignore the spin-orbital scattering effect,
which also can be responsible for the enhancement of the upper
critical field. Moreover, in the absence of spin-orbital
scattering, the transition from normal to superconducting state at
low temperatures should turn into a first-order phase transition.
\cite{Maki} Although we do not fully agree with the analysis of Ref. \onlinecite{Singleton}, we independently agree that there is a first order transition below 4 K in \ET. Even if this first order transition were not there, we suggest that spin-orbit scattering cannot be
responsible for the total enhancement of the critical field. If this
was the case, following Ref. \onlinecite{Klemm}, we calculated that the spin-orbit
scattering time would be between 0.46 and 0.62 ps, which is much less than
the total measured scattering time of about 3 ps determined from 
magnetoresistance oscillations.\cite{Singleton2}  It
is thus valuable to study the effect of the pressure on
$\hc^\|$(T) phase diagram, especially because existing data
suggests a tendency of saturation of $\hc^\|$ in \ET\ at 1.5 kbar.
\cite{Shimojo}

At ambient pressure,  it is often claimed that \ET\ is very
anisotropic. Although the anisotropy ratio $\gamma $
=$\frac{\hc^\|}{\hc^{\bot }}$ is only about 6, \cite{zuo} the
anisotropy of the London penetration depths is $\simeq$
160-330.\cite{mansky} The reason why the anisotropy determined by
the critical fields is misleading is that the mechanism that
limits the superconductivity when the applied field is parallel to
the layers is not related to the coherence length, because this
critical field is Pauli limited. Hence, the parallel and
perpendicular critical fields cannot be used to find the ratio of
the parallel and perpendicular coherence lengths, as is common
with other anisotropic superconductors.  Nevertheless, the
$\hc(\theta )$ diagram for \ET\ fits the Lawrence-Doniach 2D model
of weakly coupled layered superconductors despite the Pauli
limiting.\cite{zuo, Bayindir} And, as we will show, even under
moderate pressure  we never see results completely consistent with
anisotropic 3D Ginzburg-Landau theory.\cite{Tinkham} Thus, we hope
to discover if the effect of pressure will eventually induce a
transition from one type of superconductor (2D) to another (3D).

\section{Experimental}

The samples were single crystals of \ET\ approximately 210 $\times
$ 175 $\times $ 40 $\mu $m$^{3}$. They were placed in a four turn
coil (56 AWG wire) with an inner diameter of  300 $\mu$m with
their conducting planes perpendicular the axis of the coil. To
minimize the background signal, a nonmetallic diamond anvil cell
was used with a diamond filled epoxy gasket reinforced by a Zylon
overband.\cite{tozer2} The plastic DAC freely rotated in a top-loading dilution 
refrigerator with an ID of 21.5 mm. The coil rested in the 350
$\mu$m diameter hole of the gasket that was filled with the
quasi-hydrostatic pressure medium glycerin. Ruby was used to
calibrate the pressure at the operating temperature.\cite{forman}
The TDO setup has been explained in detail elsewhere.\cite{Coffey}
The oscillating frequency of the circuit at 70 mK was 290 MHz, and
the change in frequency during the sweep of the magnetic field was
about 2 MHz, less than 1\% percent.  Fig.~\ref{fig:ParPen} shows
typical field dependences of the frequency and amplitude when the
field is applied parallel and perpendicular to the conducting
layers. The overlap of the inverse amplitude, which is a direct
measure of the dissipation in the circuit, and the frequency,
attests to the integrity of the data. We define the critical field
as the intersection point between the linear extrapolation of the
normal state and the superconducting transition. The data reported
in the present work were taken in a top loading dilution
refrigerator and an 18 T superconducting magnet system at NHMFL in
Tallahassee. The present configuration of the TDO electronics
limits the lowest achievable temperature to 70 mK.

\section{Results}

\begin{figure}[h] \begin{center}
    \includegraphics[keepaspectratio=1,width=8 cm]{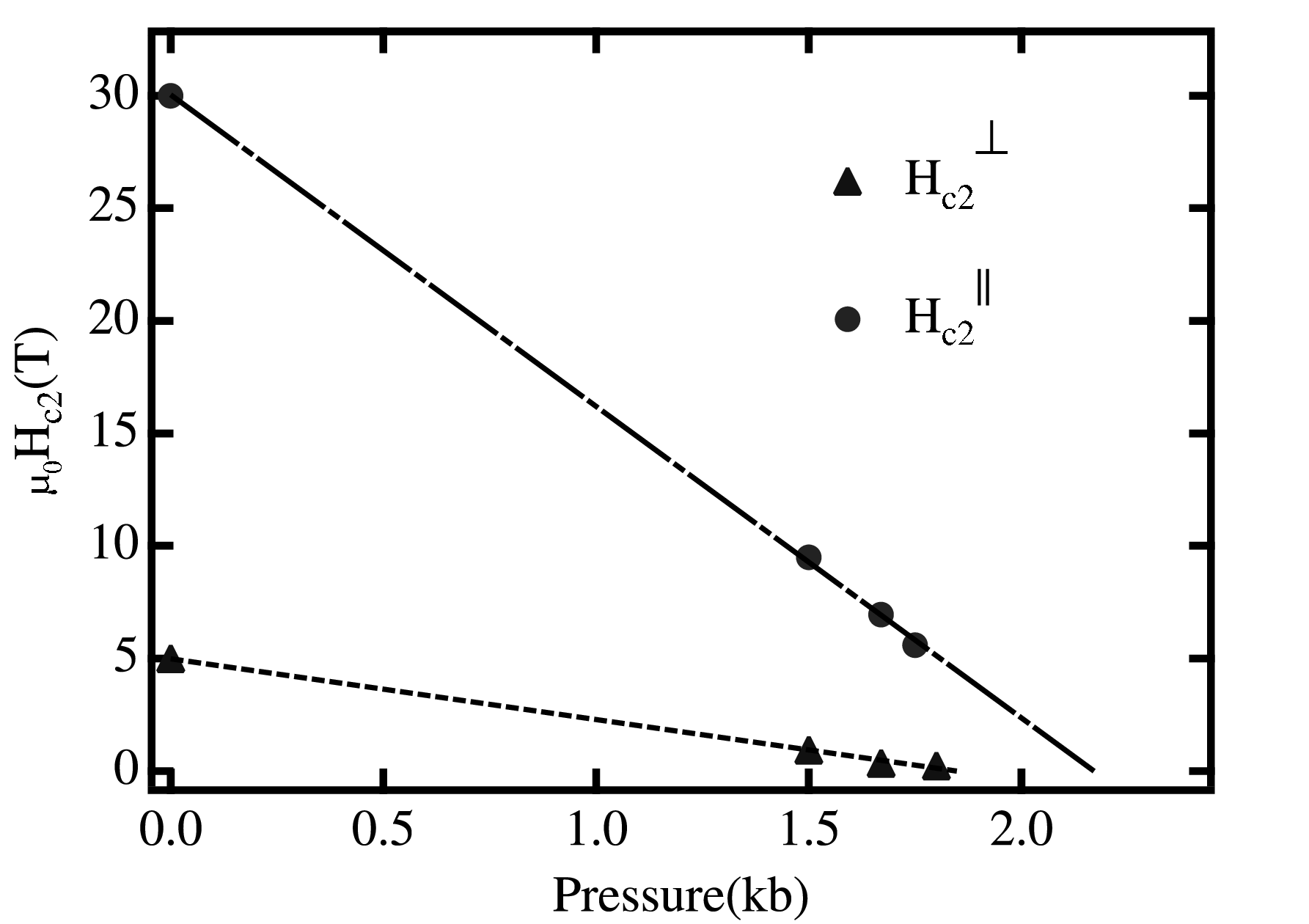}
    \caption{\label{fig:Hc2vs}Pressure dependence of parallel (circles) and perpendicular
(triangles) critical field. The error is approximately the size of the symbols. The two
ambient pressure points come from Ref. \onlinecite{zuo}. }
    \end{center}
    \end{figure}

Fig.~\ref{fig:Hc2vs} shows the critical field, both parallel and
perpendicular, for ambient pressure and three other values: 1.5, 1.67, and
1.75 kbar at T=90mK. The critical fields in the different orientations decrease linearly
as the pressure increases.
    However, the rates of change are very different for the two orientations. We found
    $d\hc^{\bot }/dP  \simeq -2.8$ Tkbar$^{-1}$ whereas
$d\hc^\|/dP \simeq -14.75$ Tkbar$^{-1}$.
Extrapolating the fitting lines, we found a critical pressure P$_c$
of about 1.8 kbar for $\hc^{\bot }$ and 2.1 kbar for $\hc^\|$. The perpendicular value
is less than half of the value reported
in Ref. \onlinecite{Caulfield} where P$_c \simeq $ 5 kbar. It is
possible that beyond 1.75 kbar the variation of $\hc$ with pressure
 is strongly nonlinear which would make our estimations invalid, although we
doubt this is the case. The suppression of superconductivity by
more than  90\% within less 1.5 kbar underlines the importance of a
careful study of the effective mass under pressure. However, the very high linear rate of change of the parallel critical field with pressure is
also striking, because while the change in the effective mass directly influences orbital effects, $\hc^\|$ is not orbitally limited.  At this
point, we only question the conclusion of Ref. \cite{Caulfield}
that the enhancement of the effective mass is directly associated
with superconductivity in \ET\ and suggest that other parameters,
such as the $V_{BCS}$ interaction term (the electron-phonon
coupling matrix element)\cite{Brooks}, the density of states
and/or the phonon characteristic energy may be very sensitive to
the applied pressure.

\begin{figure}[h] \begin{center}
    \includegraphics[keepaspectratio=1,width=8 cm]{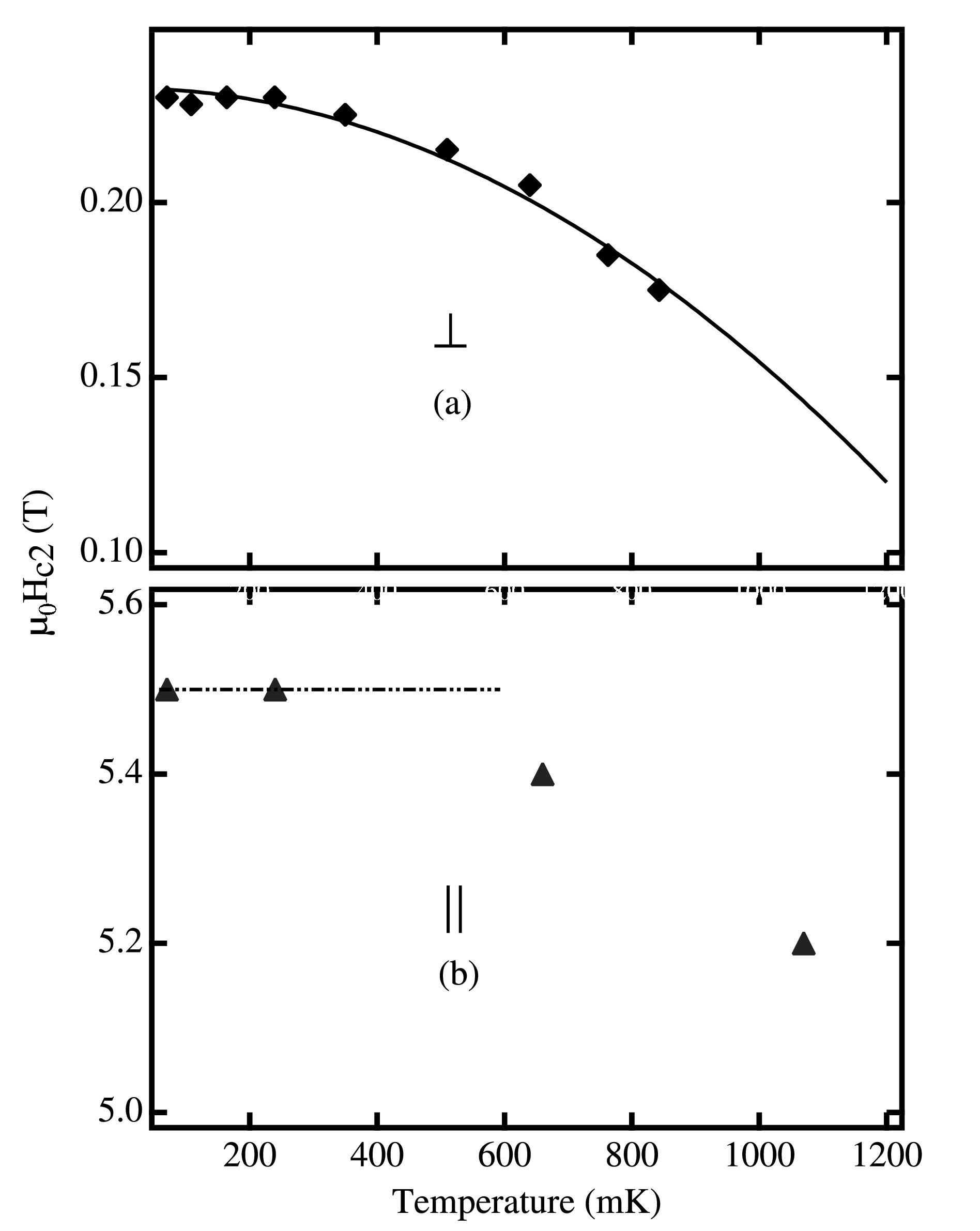}
    \caption{\label{fig:satphase}Critical fields [(a)-$\hc^{\bot}$ and (b)-$\hc^\|$] as a function of temperature at P=1.75 kbar. The continuous line in (a) is
the Ginzburg-Landau equation of the critical field at low temperature:
 $\hc$=Const.$\times (1-(T/T_c)^2).$}
    \end{center}
     \end{figure}

We measured the change in critical field with temperature at 1.75
kbar, both in the perpendicular and parallel orientation
(Fig.~\ref{fig:satphase}). Although both diagrams show a
saturation of the critical field at very low temperature, it may
not happen for the same reason as we will show later in the paper.
In the perpendicular orientation, \ET\ is orbitally limited at ambient
pressure, the orbital critical field($\approx$ 5T) being well
below the Pauli limit ($\approx$ 18 T), and we found the same
situation at 1.75 kbar. As can be seen in
Fig.~\ref{fig:satphase}(a), our experimental data falls nicely on
the theoretical Ginzburg-Landau result,  $\hc \approx
(1-(T/T_c)^2)$ albeit for only the lower half of the temperature
range. A fit and extrapolation yields  T$_c\simeq $1.75 K $\pm$
0.5 K.

As mentioned in the introduction, ambient pressure studies show
that $\hc^\|$ exceeds the Pauli paramagnetic limit $H_P^{BCS}$,
and shows no tendency of saturation as T $\to$ 0
K.\cite{Ishiguro2} In contrast, at P = 1.75 kbar we found no
change in the parallel critical field as the temperature increases
from 70 mK to 240 mK.
 Above 240 mK it drops with a negative curvature (Fig.~\ref{fig:satphase}). 
Studying how the ratio between the
measured $\hc^\|$ and the BCS Pauli limit change under
pressure, we had to sort out the very different values obtained for the
highest $\hc^\|$ at ambient pressure.  Using the upper critical field between 30 T \cite{zuo} and 35 T \cite{Singleton}, the ratio H$_{meas}$/$H_P^{BCS}$ is between 1.67
and 1.91. At 1.75 kbar, if we
conservatively estimate $T_c$ to be 2.00 K (from the perpendicular
diagram), then $H_P^{BCS}$ would be equal to 3.7 T and the ratio
H$_{meas}$/$H_P^{BCS}$ would be about 1.47.  These numbers suggest  that the effect of the pressure is to bring the parallel critical field toward the Pauli limit. 
However, TDO
measurements performed on two different sets of samples, the one
used in the present experiment and another identical to those used
in Ref \cite{Singleton}, led to a maximum critical field of 24.4 T
at 450 mK \cite{Bayindir2}. To be consistent, we compare data
obtained by the same method on the same set of samples. This data shows
that the ratio H$_{meas}$/$H_P^{BCS}$ is 1.35 at ambient pressure,
which is surprisingly less than the value of 1.47, found at 1.75 kbar.

Based on $\hc$ studied in previous experiments, one could expect either an increase or decrease in the parallel critical  field as \ET\ is subjected to pressure. If the conducting layers are decoupled and the layers are squeezed, the parallel critical field should increase as was found in single layers of aluminum.\cite{tedrow} If the insulating layers are squeezed, the parallel critical field should decrease as the the orbital limiting is enhanced  due to the increased coupling of the layers, and the increased perpendicular coherence length. It is unclear how these two phenomena will combine in a given situation, although both phenomena are tied to the energy gap and critical temperature through the density of superconducting electrons. All other parameters being equal $\hc^\|$ and $\hc^{\bot}$ will shift together. The fact that the ratio H$_{meas}$/$H_P^{BCS}$ is higher at higher pressures suggests that the conducting layers are getting thinner faster than the insulating layers, and the material is getting closer to its real Pauli limit.\cite{tedrow}  
It should also be mentioned that, up to 1.75 kbar and at temperatures 
down to 70 mK, we have seen no evidence for the FFLO phase, 
claimed to be present at ambient pressure, and the transition is 
always of second order (Fig.~\ref{fig:ParPen}(b)).

Concluding the above discussion, to understand the mechanism of
superconductivity in \ET\ it becomes very important to complete
the ambient pressure phase diagram below 500 mK, but it is
experimentally difficult to obtain magnetic
fields higher than 30-35T and temperatures below 500 mK at the same time. We have
also shown that a very low pressure, probably less than 1 kbar, would
make this experimental investigation much more facile.

However, one must always make sure that, under pressure, \ET\ does not
suffer a transition from a layered quasi 2D superconductor toward
an anisotropic 3D superconductor. If that were the case, then the orbital effects
would no longer be negligible in parallel orientation, and could even
become the dominant factor. In the case that the major effect of
the pressure is to increase the coupling between layers, this
should cause a transition from a Lawrence-Doniach type of
superconductor to an anisotropic 3-D Ginzburg-Landau one.  We can
experimentally verify this change by measuring the change in the critical
field with angle. The Ginzburg-Landau theory for an anisotropic
3-D superconductor predicts a variation of the critical with angle
after the following equation: \cite{Tinkham}

\begin{equation}
\left[\frac{\hc(\theta)cos(\theta)}{\hc^{\bot}}\right] ^2+
\left[ \frac{\hc(\theta)sin(\theta)}{\hc^{\|}}\right] ^2=1,
\end{equation}
where $\theta $ is the angle between the field and the normal to the layers.
For weakly coupled layered superconductors, Tinkham\cite{Tinkham} and then
Schneider and Schmidt\cite{Schneider} found that the angular dependence is given by:

\begin{equation}
\left|\frac{\hc(\theta)cos(\theta)}{\hc^{\bot}}\right| +
\left[ \frac{\hc(\theta)sin(\theta)}{\hc^{\|}}\right] ^2=1,
\end{equation}
which leads to a cusp-like behavior.

\begin{figure}[h] \begin{center}
   \includegraphics[keepaspectratio=1,width=8 cm]{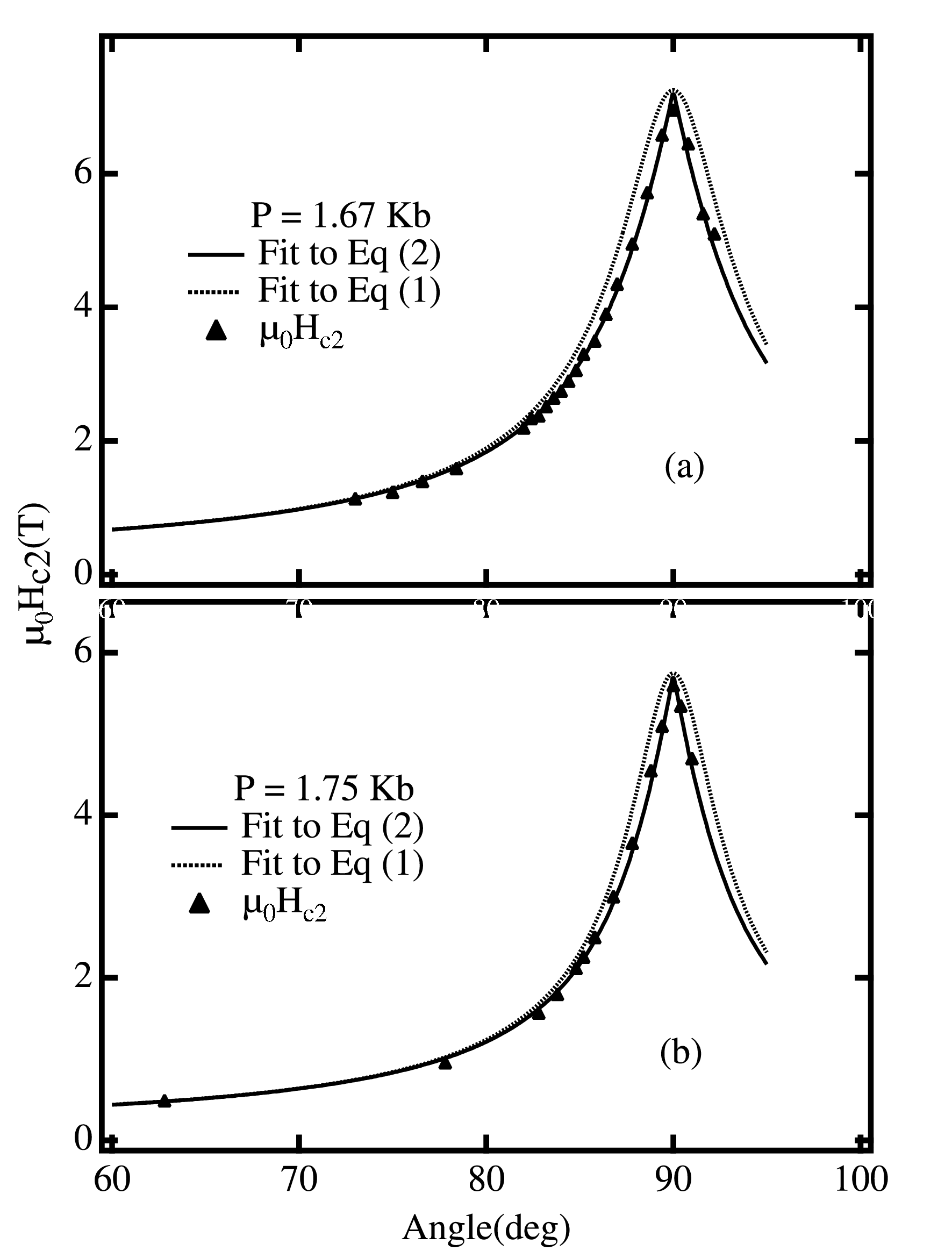}
    \caption{\label{fig:angular}Angular dependence of $\hc$ at 1.67 kbar (a), and 1.75
kbar (b). For both graphs, the continuous line represents a fit with
Lawrence-Doniach equation and the dotted curve is a fit to the anisotropic 3D
Ginzburg-Landau equation.}
    \end{center}
     \end{figure}

We have determined the $\hc$($\theta $) diagram for P=1.67 and
1.75 kbar at T=70 mK. The experimental result along with the fits
by Eq.(1) and Eq.(2), are plotted in Fig.~\ref{fig:angular}. The
cusp-like feature observed experimentally at $\theta
$=90${{}^{\circ }}$ is the indication that Eq.(2) is a better fit
up to 1.75 kbar. Therefore, we confirm experimentally that up to
P=1.75 kbar there is no evidence for moving toward a more 3-D (or
less 2-D) superconductor. We also found an enhancement of the
anisotropy in critical field $\gamma $=$\frac{\hc^\|}{\hc^{\bot
}}$
 from $\simeq$~6 at ambient
pressure to $\gamma \simeq ~$23 at P=1.75 kbar, but we attribute
this increase in the apparent anisotropy to the different
mechanisms that affect the critical fields in each orientation
rather than to an enhancement of the 2-D character of the \ET, as
discussed in the introduction. In reference to our comment earlier that $\hc^\|$ at 1.75 kbar is Pauli limited, the cusp like character of the angular dependence is further evidence that the orbital effects are suppressed when the sample is at the parallel orientation.  If there was significant transport through the layers, the angular dependence would have a rounded top near $90^o$

\begin{figure}[h] \begin{center}
    \includegraphics[keepaspectratio=1,width=8 cm]{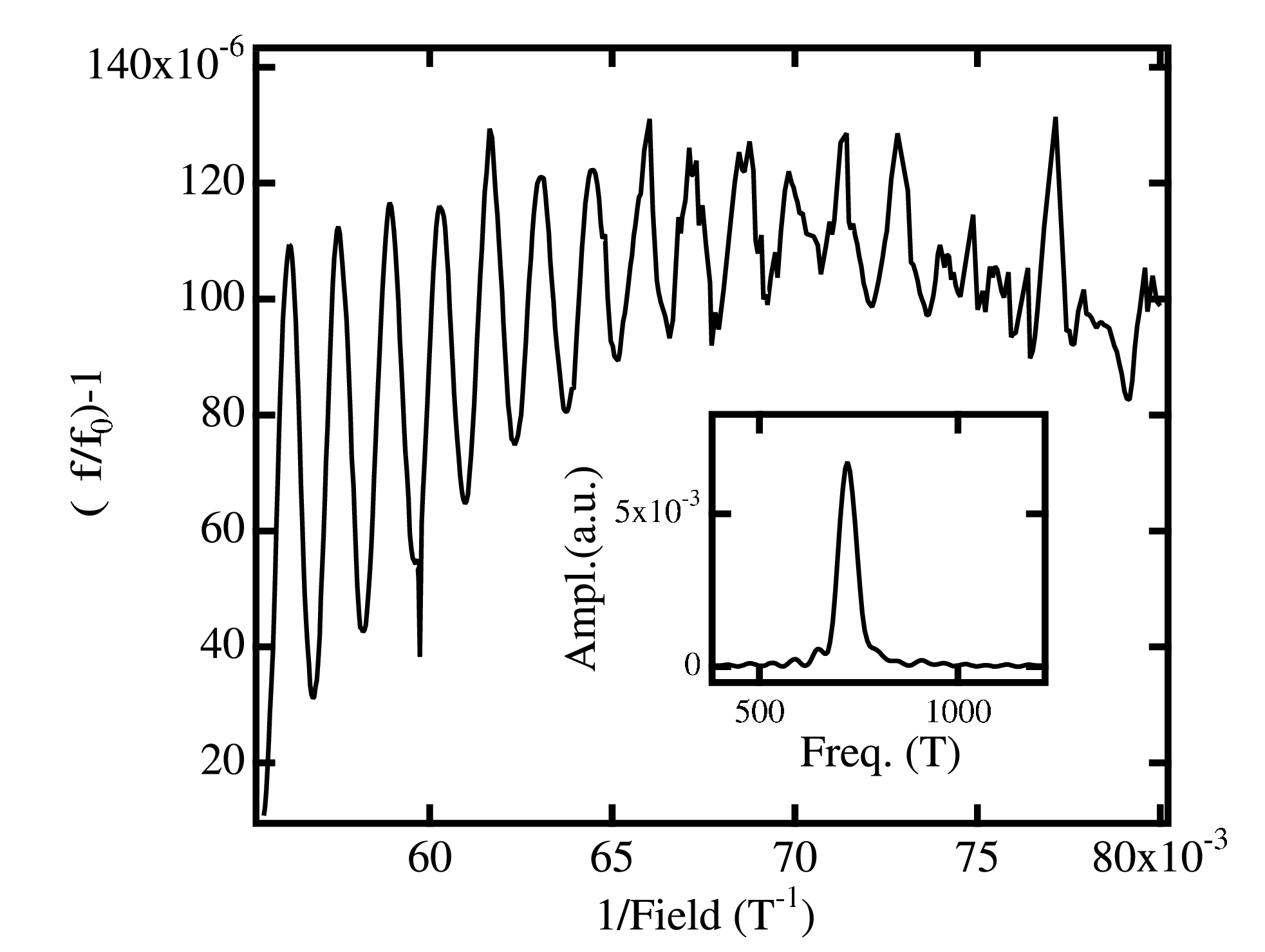}
    \caption{\label{fig:sdh}Magnetoresistance oscillations of \ET\ at P=1.67 kbar and T=90 mK. The inset shows the FFT of the oscillations.}
    \end{center} \end{figure}

Beyond the superconducting transition, the change in frequency (and
amplitude) of the TDO is due to the resistivity of the normal state, and at
higher fields we have measured the Shubnikov-de Haas oscillations in
magnetoresistance as shown in Fig.~\ref{fig:sdh}. Our limit of 18 tesla did not
allow for a careful analysis of the oscillation frequency with pressure
and temperature, but we found an increase in the frequency of $\alpha $%
-orbit (F$_{\alpha }$) from 694.1 T at 1.5 kbar to 703.0 T at 1.67 kbar
(T=90 mK), while the ambient value (F$_0$) is about 595 T. (At 1.75 kbar an experimental
error prevented us from seeing the SdH oscillations.) The
ratio $\frac{F_{\alpha }}{F_0}$ is therefore, 1.17 at 1.5 kbar and
1.18 at 1.67 kbar.  The linear
increase of the frequency of oscillations with pressure is due to the
change in size of the unit cell and the inverse effect on the Brillouin zone.
\cite{Caulfield, Brooks,Ivanov}
\section{Conclusions}

\smallskip In summary, we have proven that the combination of
the TDO technique and the nonmetallic pressure cell can provide a
very useful tool in the study of superconductivity.

Measurements on \ET\ revealed that the pressure strongly affects
the critical field, by more than 90\% within 1.5 kbar, both in the
perpendicular and parallel orientations. The superconductivity is
suppressed at a much higher rate in the parallel direction, although
in this case the orbital critical field, directly proportional to
the effective mass, is not the limiting factor. 

At 1.75 kbar, we found a clear change in the behavior of the
parallel critical field with temperature, from the ambient
pressure phase diagram. The value of $\hc^\|$ still exceeds the
BCS Pauli limit, but the lack of consistent data at ambient
pressure confuses the issue of how the ratio
H$_{meas}$/$H_P^{BCS}$ evolves with pressure. According to our
experimental evidence, up to 1.75 kbar,  \ET\, is still well
described by the 2-D Lawrence-Doniach model for layered
superconductors.

The frequency of magnetoresistance oscillations increases with
pressure at a higher rate than previously reported in literature
Ref.\cite{Caulfield}. In an effort to better understand the role
played by different physical quantities (e.g. the effective mass,
V$_{\text{BCS}}$, transfer integral, spin-orbit scattering rate)
we are pursuing larger fields, lower temperatures, and higher
pressures. Exploring the very low pressure gap in our data would
be very useful as well.

The authors would like to thank Tim Murphy and Eric Palm for their help during this
experiment, J. Singleton for helpful discussions and R. Desilets and J. Farrell for
careful machining. This work was supported by the NSF Cooperative Agreement No.
DMR-0084173 at the NHMFL and, in particular, the NHMFL In House Research Program.

\end{document}